\documentclass[12pt]{iopart}

\usepackage{braket}
\usepackage{graphicx}
\begin{document}

\title{An improved bound for strong unitary uncertainty relations with refined sequence}

\author{Jing Li$^{1}$, Sujuan Zhang$^{1,2}$, Lu Liu$^{1}$ and Chen-Ming Bai$^{1}$}

\address{$^{1}$Department of Mathematics and Physics, Shijiazhuang Tiedao University, Shijiazhuang, 050043, China}
\address{$^{2}$Postdoctoral Research Station of Mathematics, Hebei Normal University, Shijiazhuang, Hebei, 050024, China}
\ead{zhangsj@stdu.edu.cn}
\vspace{10pt}

\begin{abstract}
We derive the lower bound of uncertainty relations of two unitary operators for a class of states based on the geometric-arithmetic inequality and Cauchy-Schwarz inequality. Furthermore, we propose a set of uncertainty relations for three unitary operators. Compared to the known bound introduced in Phys.Rev.A.100,022116(2019), the unitary uncertainty relations bound with our method is tighter, to a certain extent. Meanwhile, some examples are given in the paper to illustrate our conclusions.

\noindent{\ \textbf{Keywords}: unitary operator, uncertainty relations, lower bound\/}
\end{abstract}

\section{Introduction}

Uncertainty relations play an important role in quantum mechanics, which reveal the difference between the classic world and the quantum world. It has strong application prospects, such as entanglement detection \cite{Hofmann2003, G¨¹hne2004}, quantum metrology \cite{Giovannetti2004}-\cite{Braunstein1996}, quantum cryptography \cite{Ekert1991}-\cite{Bai}, signal processing \cite{Candes2006}, quantum speed limit \cite{Pires2016}, and so on. In 1927, the uncertainty principle was first proposed by Heisenberg \cite{Heisenberg1927}, which later formulated by Kennard as \cite{Kennard1927}:
\begin{equation}
\Delta \hat{x}\Delta \hat{p}\geq \frac{1}{2},
\end{equation}
where $\hat{x}$ and $\hat{p}$ are the position and momentum observables respectively. Subsequently, Robertson \cite{Robertson1929} and Schr\"{o}dinger \cite{Schrodinger1930} generalized this uncertainty relations to any two non-commuting observables $A$ and $B$ and a fixed state $|\psi\rangle$,
\begin{equation}
\Delta A\Delta B\geq \frac{1}{2}|\langle\psi|[A, B]|\psi\rangle|,
\end{equation}
where $\Delta A=\sqrt{\langle A^{2}\rangle-\langle A\rangle^{2}}$,  $\Delta B=\sqrt{\langle B^{2}\rangle-\langle B\rangle^{2}}$, and $\langle O\rangle=\langle\psi|O|\psi\rangle$ is the average for an observables $O$ in the state $|\psi\rangle$.
Recently, many scholars did a lot of research on the uncertainty relations based on entropy and variance.

With the development of quantum information theory, it is natural to characterize the uncertainty via information entropy. The entropy \cite{Hertz2018}-\cite{Dong} uncertainty relations for any pair of observables was given by Deutsch \cite{Deutsch1983}. An improvement of Deutsch's entropy uncertainty relations was subsequently conjectured by Kraus \cite{Kraus1987} and proved by Maassen and Uffink \cite{Maassen1988}.

On the other hand, there are also many achievements on the uncertainty relations based on variance. Massar and Spindel proved the uncertainty relations for two unitary operators that obey the commutation relation $UV=e^{i\phi}VU$, which applies to constrain for a quantum state can be localized simultaneously in two mutually unbiased bases related by a discrete Fourier transform \cite{Massar2008}.
Later, some further uncertainty relations related by discrete Fourier transform for unitary operators were presented in \cite{Klimov2009}-\cite{Marchiolli2013}. In 2014, Maccone and Arun \cite{Maccone2014} presented two stronger uncertainty relations connected to the sum of the variances, as long as the two observables are incompatible with the system state, the lower bound is guaranteed to be nontrivial. Soon after, Li and Qiao \cite{Li2015} introduced a new uncertainty relations which may propose a complete trade-off relations for variances of observables in pure and mixed quantum systems. Their bounds are independent of the quantum state and are not affected by the problem of expecting zero. Bagchi and Pati \cite{Bagchi2016} put forward the sum form of variance-based uncertainty relations for two general unitary operators before long, which was tested by experimentally. Then Mondal \emph{et al.}\cite{Mondal2017} derived tighter upper and lower bounds for both the product and sum forms of the variance-based uncertainty relations. Later on, Sharma \emph{et al.}\cite{Sharma2018} proposed the mean-deviation-based uncertainty relations, in both state-dependent and state-independent forms for a general set of deviation measures. Following Xiao \emph{et al.}'s method \cite{Xiao2016} for a sequence of \textquotedblleft fine-grained\textquotedblright inequalities, Yu \emph{et al.}\cite{Yu2019} used this method to derive variance-based unitary uncertainty relations in the product form for two and three unitary operators in all quantum systems, and their uncertainty bounds are tighter than the bound in \cite{Bong2018}. However, the uncertainty bounds for unitary operators are not tight enough. Hence finding a tighter lower bound of the uncertainty relations is a problem worth studying.

In this paper, we improve the lower bounds for strong unitary uncertainty relations. For two unitary operators, we obtain a lower bound that is tighter than Yu \emph{et al.}\cite{Yu2019} for a class of states. Meanwhile, we find that the partition of descending sequence given in the uncertainty relations for two unitary operators \cite{Yu2019} is not detailed enough. Hence we improve the descending sequence to be finer. And
the lower bound of uncertainty relations for three unitary operators is deduced by using the improved descending sequence.

The structure of this paper is as follows. In Sec.II we first review some of the concepts and knowledge of uncertainty relations. Then for two unitary operators, we propose a lower bound of the uncertainty relations for a class of states. In additions, a more dense descending sequence is constructed based on Cauchy-Schwarz inequality. In Sec.III according to the variable separation method, we establish the relationship of descending sequence between two unitary operators and three unitary operators, and then obtain product-form variance-based unitary uncertainty relations for three unitary operators.

\section{Uncertainty relations for two unitary operators}

Let $A$ and $B$ be two arbitrary finite-dimensional unitary operators defined in a Hilbert space. The variances of operators $A$ and $B$ in the state $\ket{\psi}$ are defined as
\begin{eqnarray}
  \Delta A^2 =\langle(A-\langle A\rangle)^{\dag}(A-\langle A\rangle)\rangle =\langle\psi|\bar{A}^{\dag}A|\psi\rangle,\\
  \Delta B^2=\langle(B-\langle B\rangle)^{\dag}(B-\langle B\rangle)\rangle =\langle\psi|\bar{B}^{\dag}B|\psi\rangle,
\end{eqnarray}
where $\bar{A}=A-\langle A\rangle$, $\bar{B}=B-\langle B\rangle$.
It is easy to see $0\leq\Delta A^2\leq1$, $0\leq \Delta B^2\leq1$.

By choosing  a computational basis $\{\ket{\psi_1}, \cdots, \ket{\psi_n}\}$, the state $\ket f=\bar{A}\ket{\psi}$ can be writted as $\ket f=\sum\limits_{i=1}^n\alpha_i\ket{\psi_i}$. Similarly we have the state $\ket g=\bar{B}\ket{\psi}=\sum\limits_{j=1}^n\beta_j\ket{\psi_j}$.
Let $\Delta A^{2}=|\vec{X}|^2$ (resp. $\Delta B^{2}=|\vec{Y}|^2$) for the (nonnegative) real vectors $\vec{X}=(x_1, x_2, \cdots, x_n)$ (resp. $\vec{Y}=(y_1, y_2, \cdots, y_n)$), where $ x_i=|\alpha_i|$, $y_j=|\beta_j|$.
Then the product of the variances 
can be rewritten as $\Delta A^2\Delta B^2=|\vec{X}|^2|\vec{Y}|^2=\sum\limits_{i,j}^{n}x_i^2y_j^2$ \cite{Yu2019}.

For the lower bounds of uncertainty relations of two unitary operators, many conclusions have been given. On the basis of previous conclusions, we give the following theorem.

\emph{Theorem 1.} Let $H$ be an $n$-dimensional $(n\geq 3)$ Hilbert space, $\rho$ is a fixed quantum state, $A$ and $B$ are two unitary operators. The product of the variances of $A$ and $B$ satisfies
the following uncertainty relations
\begin{equation}
\Delta A^2\Delta B^2\geq I_{1}',
\end{equation}
where
\begin{equation}
I_{1}'=\sum_{i=1}^{n}x_{i}^{2}y_{i}^{2}+\sum_{j\neq 1\atop i\neq j}^{n}x_{i}^{2}y_{j}^{2}+y_{1}^{2}\sum_{i=4}^{n}x_{i}^{2}+2y_{1}^{2}x_{2}x_{3},
\end{equation}
and the equality holds if and only if $x_{2}=x_{3}$.

\emph{Proof.} From the above we know that
\begin{eqnarray*}
  \Delta A^{2}\Delta B^{2} &=& \sum_{i,j}^{n}x_{i}^{2}y_{j}^{2}\\
  &=& \sum_{i=1}^{n}x_{i}^{2}y_{i}^{2}+\sum_{j\neq 1\atop i\neq j}^{n}x_{i}^{2}y_{j}^{2}+y_{1}^{2}\sum_{i=2}^{n}x_{i}^{2}\\
  &=&\geq \sum_{i=1}^{n}x_{i}^{2}y_{i}^{2}+\sum_{j\neq 1\atop i\neq j}^{n}x_{i}^{2}y_{j}^{2}+y_{1}^{2}\sum_{i=4}^{n}x_{i}^{2}+2y_{1}^{2}x_{2}x_{3} \\
  &=& I_{1}',
\end{eqnarray*}
where the inequality is due to the Cauchy-Schwarz inequality.

Without loss of generality, let us discuss the case of $n=3$, we have
\begin{eqnarray*}
\Delta A^{2}\Delta B^{2}&=\sum_{i,j}^{3}x_{i}^{2}y_{j}^{2}\\
&=\sum_{i=1}^{3}x_{i}^{2}y_{i}^{2}+\sum_{j\neq 1\atop i\neq j}^{3}x_{i}^{2}y_{j}^{2}+y_{1}^{2}\sum_{i=2}^{3}x_{i}^{2}\\
&\geq \sum_{i=1}^{3}x_{i}^{2}y_{i}^{2}+\sum_{j\neq 1\atop i\neq j}^{3}x_{i}^{2}y_{j}^{2}+2y_{1}^{2}x_{2}x_{3}\\
&=I_{1}'.
\end{eqnarray*}

Recently, Yu \emph{et al.} derived a strong the variance-based uncertainty relations for two unitary operators \cite{Yu2019}. In this paper, they defined $(1\leq d\leq n)$
\begin{eqnarray}
I_{d}=\sum\limits_{1\leq i\leq n}x_{i}^{2}y_{i}^{2}+\sum\limits_{1\leq i<j\leq n\atop d<j}(x_{i}^{2}y_{j}^{2}+x_{j}^{2}y_{i}^{2})+\sum\limits_{1\leq i<j\leq d}2x_{i}y_{i}x_{j}y_{j}.
\end{eqnarray}
According to the recursive formula $I_{d+1}-I_{d}=-\sum\limits_{i=1}^{d}(x_{i}y_{d+1}+y_{i}x_{d+1})^{2}\leq 0$, they obtained the descending sequence
\begin{equation}
I_{1}\geq I_{2}\geq \cdots \geq I_{n-1}\geq I_{n}.
\end{equation}
The result indicates that $\Delta A^{2}\Delta B^{2}\geq I_{d}$, where $I_{2}$ is the optimal bound in this case.

However, for a class of states, we can obtain the lower bound  $I_{1}'$ of two unitary operators from Theorem 1, which is tighter than $I_{2}$.

Next, in order to illustrate the superiority of our bound $I_{1}'$ , the following examples will be given.

\emph{Example} 1. Suppose the pure states $|\varphi\rangle=\cos\theta|0\rangle+\sin\theta|2\rangle$ on a Hilbert space, and $A$, $B$ are two unitary operators:

\begin{equation}
A=\left(
\begin{array}{ccc}
1 & 0 & 0 \\
0 & e^{\frac{2\pi i}{3}} & 0 \\
0 & 0 & e^{\frac{4\pi i}{3}}
\end{array}
\right),
B=\left(
\begin{array}{ccc}
0 & 0 & 1 \\
1 & 0 & 0 \\
0 & 1 & 0
\end{array}
\right).
\end{equation}

Their associated real vectors $\vec{X}=(x_{1}, x_{2}, x_{3})$, $\vec{Y}=(y_{1}, y_{2}, y_{3})$ are given by
\begin{equation}
x_{1}=|(1-e^{-\frac{2\pi i}{3}})\sin^{2}\theta \cos\theta|, x_{2}=0,x_{3}=|(e^{-\frac{2\pi i}{3}}-1)\sin\theta \cos^{2}\theta|,
\end{equation}
and
\begin{equation}
y_{1}=|\sin^{3}\theta|, y_{2}=|\cos\theta|,y_{3}=|-\sin^{2}\theta \cos\theta|.
\end{equation}

In this example, we find that the green dotted curve is always above the horizontal axis in Fig.1, which means that $I_{1}'-I_{2}$ is always greater than zero regardless of the value of $\theta$. Therefore, for this class of states, our bound $I_{1}'$ is tighter than the bound in \cite{Yu2019}.

\emph{Example} 2. Let us consider the pure states $|\varphi\rangle=\frac{\sqrt{2}}{2}\cos\theta|0\rangle+\frac{\sqrt{2}}{2}\cos\theta|1\rangle+\sin\theta|2\rangle$ in a Hilbert space. Here the two unitary operators $A$ and $B$ are the unitary operators in example 1.

Similarly, their associated real vectors $\vec{X}$ and $\vec{Y}$ also can be calculated.
We observe that $I_{1}'-I_{2}$ has both positive and negative values in Fig.2 , which is different from example 1. For $I_{1}'-I_{2}> 0$, our bound $I_{1}'$ is always greater than $I_{2}$. For $I_{1}'-I_{2}< 0$, the curve $I_{1}'$ is always below the curve $I_{2}$. As in the subgraph in Fig.2, when $\theta\in[4,4.6]$, the curve $I_{1}'$ is always above $I_{2}$, that is, for this class of states, our bound $I_{1}'$ is tighter than bound in \cite{Yu2019}.

\begin{figure}[h]
\begin{center}
\centerline{\includegraphics[width=0.6\textwidth]{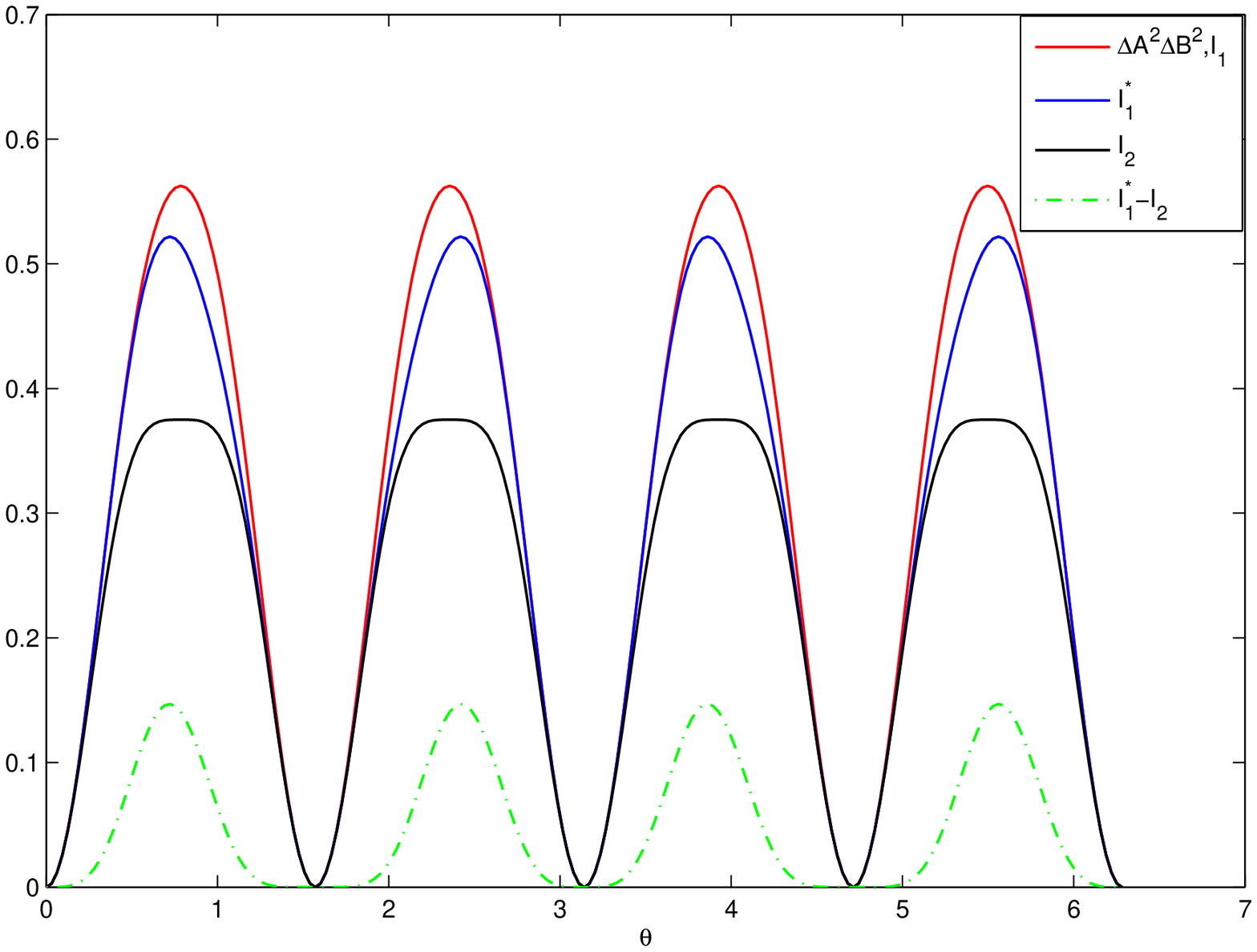}}
\caption{\textbf{Comparison of our bound with Yu \emph{\textbf{et al}}.'s bound for pure state.} The red (upper) and black curves represent $\Delta A^2\Delta B^2$ ($I_{1}$) and Yu \etal.'s bound $I_2$ respectively.
    The blue curve represents our bound $I_{1}'$, the green dotted curve represents the condition of $I_{1}'-I_{2}$.}
\label{fig1}
\end{center}
\end{figure}

\begin{figure}[h]
\begin{center}
\centerline{\includegraphics[width=0.6\textwidth]{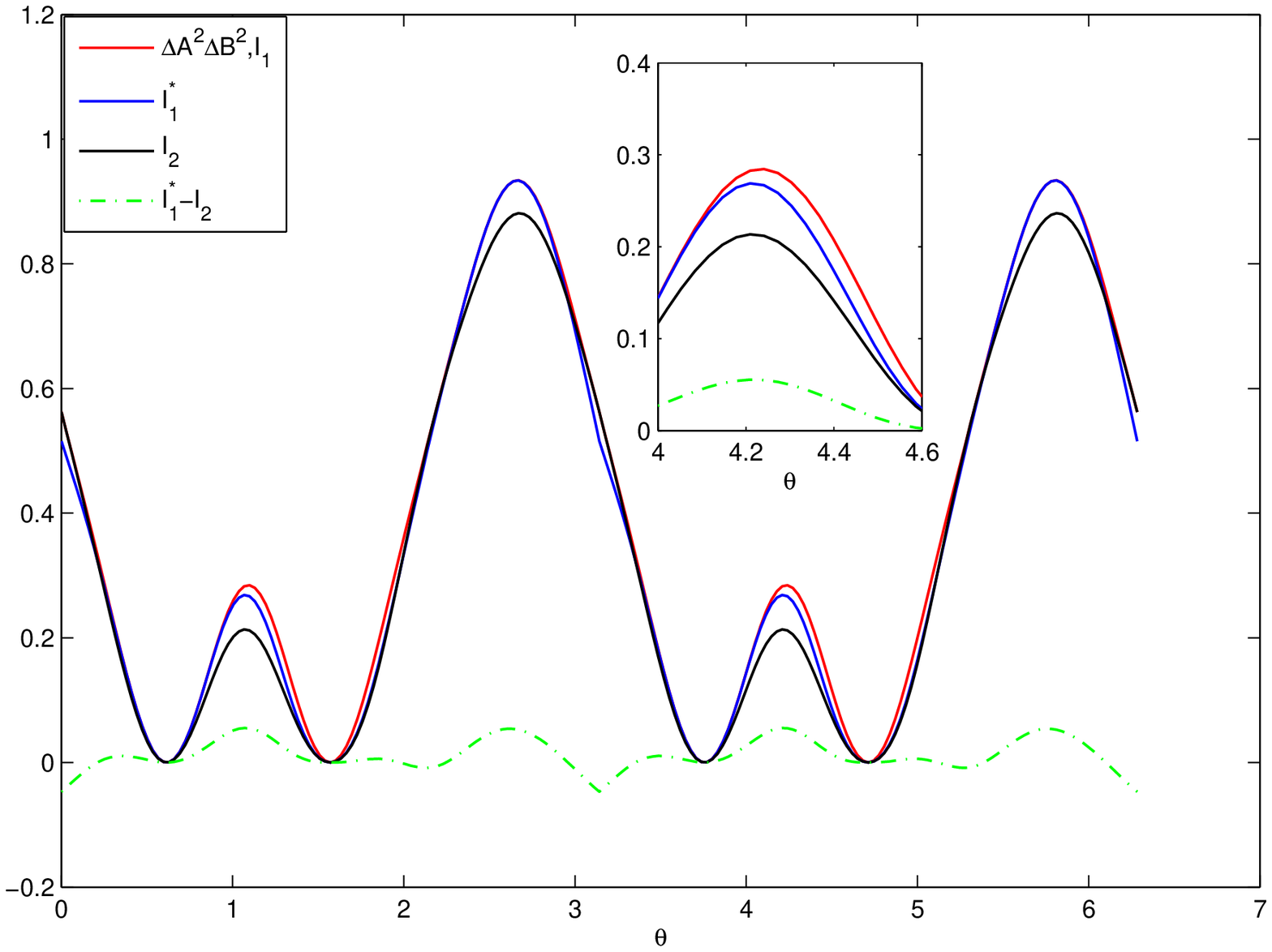}}
\caption{\textbf{Comparison of our bound with Yu \emph{\textbf{et al}}.'s bound for pure state.} The red (upper) and black curves represent $\Delta A^2\Delta B^2$ ($I_{1}$) and Yu \etal.'s bound $I_2$ respectively,
    the blue curve represents our bound $I_{1}'$, the green dotted curve represents the condition of $I_{1}'-I_{2}$.}
\label{fig2}
\end{center}
\end{figure}

For two unitary operators, we can also strengthen the bound using the symmetry of $S_{N}$, which acts on the set $\{1, 2,\cdots, N\}$. For example, when $n=3$, let $\pi_{1}, \pi_{2}\in S_{N}$ are two any permutations, we define
\begin{equation}
(\pi_{1}, \pi_{2})I_{1}'=\sum_{1\leq i\leq 3}x_{\pi_{1}(i)}^{2}y_{\pi_{2}(i)}^{2}+\sum_{j\neq 1\atop i\neq j}^{3}x_{\pi_{1}(i)}^{2}y_{\pi_{2}(j)}^{2}+2y_{\pi_{1}(1)}^{2}x_{\pi_{2}(2)}x_{\pi_{2}(3)},
\end{equation}
then
\begin{equation}
\Delta A^{2}\Delta B^{2}\geq \max\limits_{\pi_{1}, \pi_{2}\epsilon S_{N}}(\pi_{1}, \pi_{2})I_{1}'.
\end{equation}
Similarly, for $n>3$, we can define $(\pi_{1}, \pi_{2})I_{1}'$, then obtain the result $\Delta A^{2}\Delta B^{2}\geq \max\limits_{\pi_{1}, \pi_{2}\epsilon S_{N}}(\pi_{1}, \pi_{2})I_{1}'$.

On the other hand, we can find that the descending sequence (8) is not enough tight after analysis. Hence next we will improve the descending sequence.

Now we refine the descending sequence by introducing Cauchy-Schwarz inequality.
For each $p\geq 3$ and $q=1,2,\cdots,(p-1)$, we define $(p>q)$
\begin{equation}
S_{pq}=-\sum_{j=2\atop j>i}^{p-1}\sum_{i=1}^{j-1}(x_{j}y_{i}-x_{i}y_{j})^{2}-\sum_{m=1}^{q}(x_{p}y_{m}-x_{m}y_{p})^{2}
+\sum_{i,j}^{n}x_{i}^{2}y_{j}^{2}.
\end{equation}
In particular, for $p=1, q=0$, we have
$S_{10}=\sum\limits_{i,j}^{n}x_{i}^{2}y_{j}^{2}$.
For $p=2, q=1$, we have
$S_{21}=\sum\limits_{1\leq i\leq n }x_{i}^{2}y_{i}^{2}+\sum\limits_{1\leq i<j\leq n\atop 2<j}(x_{i}^{2}y_{j}^{2}+x_{j}^{2}y_{i}^{2})+2x_{1}y_{1}x_{2}y_{2}$.

The quantities $S_{pq}$ can be visualized by lattice dots within an $n\times n$ square as follows. In Fig.3,
the black dot at $i$th column and $j$th row represents $x_i^2y_j^2$. Based on $S_{10}$, $S_{21}$ is derived by the Cauchy-Schwarz inequality for $x_{2}^{2}y_{1}^{2}$ and $x_{1}^{2}y_{2}^{2}$. $S_{31}$ is derived based on $S_{21}$ by the Cauchy-Schwarz inequality for $x_{3}^{2}y_{1}^{2}$ and $x_{1}^{2}y_{3}^{2}$. $S_{32}$ is derived based on $S_{31}$ by the Cauchy-Schwarz inequality for $x_{3}^{2}y_{2}^{2}$ and $x_{2}^{2}y_{3}^{2}$, and so on. Similarly we can get each quantity $S_{pq}$.

\begin{figure}[h]
\begin{center}
\centerline{\includegraphics[width=0.4\textwidth]%
  {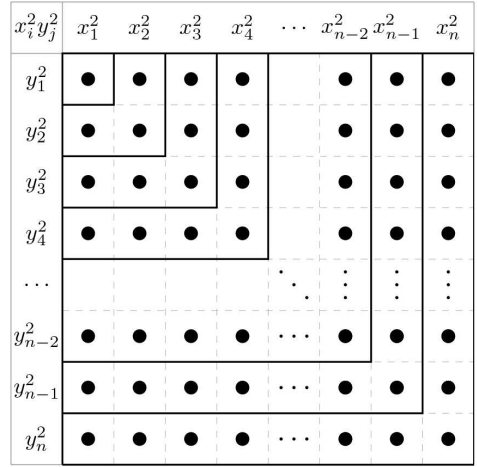}}
\caption{\textbf{Diagram for the \textbf{$S_{pq}$ ($p=1,2,\cdots,n; q=1,2,\cdots,(p-1)(p>q)$)}} The black $(i, j)$-dot represents $x_i^2y_j^2$. $S_{pq}$ is derived based on $S_{p(q-1)}$ by the Cauchy-Schwarz inequality for $x_{p}^{2}y_{q}^{2}$ and $x_{q}^{2}y_{p}^{2}$. $S_{p1}$ is derived based on $S_{(p-1)(p-2)}$ by the Cauchy-Schwarz inequality for $x_{p}^{2}y_{1}^{2}$ and $x_{1}^{2}y_{p}^{2}$.}
\end{center}
\end{figure}

It is easily seen that
\begin{eqnarray*}
S_{21}-S_{10}=-(x_{2}y_{1}-x_{1}y_{2})^{2}\leq 0,\\
S_{pq}-S_{p(q-1)}=-(x_{p}y_{q}-y_{q}x_{p})^{2}\leq0,\\
S_{p1}-S_{(p-1)(p-2)}=-(x_{p}y_{1}-x_{1}y_{p})^{2}\leq0.
\end{eqnarray*}
For this reason, a descending sequence involving these $S_{pq}$s can be constructed as
\begin{equation}
  S_{10}\geq S_{21}\geq S_{31}\geq S_{32}\geq S_{41}\geq\cdots
\geq S_{n1}\geq S_{n2}\geq S_{n3}\geq\cdot\cdot\cdot\geq S_{n(n-1)}.
\end{equation}

Next, we will briefly illustrate that this descending sequence is hold.
According to the quantities $S_{pq}$ given by equation (14), we have $S_{10}=\sum\limits_{i,j}^{n}x_{i}^{2}y_{j}^{2}=\Delta A^{2}\Delta B^{2}$. Then from the recursive relation given above, the following recursive terms can be obtained
\begin{eqnarray*}
S_{21}-S_{10}=-(x_{2}y_{1}-x_{1}y_{2})^{2}\leq 0,\\
S_{31}-S_{21}=-(x_{3}y_{1}-x_{1}y_{3})^{2}\leq 0,\\
S_{32}-S_{31}=-(x_{3}y_{2}-x_{2}y_{3})^{2}\leq 0,\\
S_{41}-S_{32}=-(x_{4}y_{1}-x_{1}y_{4})^{2}\leq 0,\\
\qquad \vdots\\
S_{n1}-S_{(n-1)(n-2)}=-(x_{n}y_{1}-x_{1}y_{n})^{2}\leq 0,\\
\qquad \vdots\\
S_{n(n-1)}-S_{n(n-2)}=-(x_{n}y_{n-1}-x_{n-1}y_{n})^{2}\leq 0.
\end{eqnarray*}
Therefore we obtain the descending sequence as follows:
$$S_{10}\geq S_{21}\geq S_{31}\geq S_{32}\geq S_{41}\geq\cdots\geq S_{n1}\geq S_{n2}\geq S_{n3}\geq\cdots\geq S_{n(n-1)}.$$

We can easily find that the descending sequence (15) is \textquotedblleft fine-grained\textquotedblright compared with the sequence (8) in \cite{Yu2019}. For the quantities $S_{pq}$, when $q=p-1$, we have
\begin{eqnarray}
S_{p(p-1)}&=\sum\limits_{i,j}^{n}x_{i}^{2}y_{j}^{2}-\sum\limits_{j=2}^{p-1}
\sum\limits_{i=1}^{j-1}(x_{j}y_{i}-x_{i}y_{j})^{2}-\sum\limits_{m=1}^{p-1}
(x_{p}y_{m}-x_{m}y_{p})^{2}\nonumber\\
&=\sum\limits_{1\leq i\leq n}x_{i}^{2}y_{i}^{2}+\sum\limits_{1\leq i<j\leq n\atop p<j}(x_{i}^{2}y_{j}^{2}+x_{j}^{2}y_{i}^{2})+\sum\limits_{1\leq i<j\leq p}2x_{i}y_{i}x_{j}y_{j}=I_{p}.
\end{eqnarray}
At the same time, it is easily to compute $S_{10}=I_{1}$ and $S_{21}=I_{2}$. Then we have
\begin{eqnarray}
S_{10}&=I_{1}\geq S_{21}=I_{2}\geq S_{31}\geq S_{32}=I_{3}\geq S_{41}\geq S_{42}\nonumber\\
&\geq S_{43}=I_{4}\geq \cdots \geq S_{n1}\geq S_{n2}\geq \cdots \geq S_{n(n-1)}=I_{n}.
\end{eqnarray}

This means that our sequence is refined compared with the sequence (8). Meanwhile, the descending sequence of $S_{pq}$ also lays a foundation for us to study the uncertainty relations of three unitary operators later.

To see what we're doing more intuitively, let us take an example and illustrate with figures we draw.

\emph{Example} 3. We take this class of pure states $|\psi\rangle=\frac{\sqrt{d-1}}{d-1}\cos\theta\sum\limits_{a=0}^{d-2}|a\rangle-\sin\theta|d-1\rangle$ on the $d$-dimensional Hilbert space $(d\geq 3)$. Suppose $A$ and $B$ are the following unitary operators:

\begin{equation}
A=\left(
\begin{array}{ccccc}
1 & 0 & 0 & \cdots & 0\\
0 & \omega & 0 & \cdots & 0\\
0 & 0 & \omega^2 & \cdots & 0\\
\vdots & \vdots & \vdots & \ddots & \vdots\\
0 & 0 & 0 & \cdots & \omega^{d-1}
\end{array}
\right),
B=\left(
\begin{array}{ccccc}
0 & 0 & \cdots & 0 & 1\\
1 & 0 & \cdots & 0 & 0\\
0 & 1 & \cdots & 0 & 0\\
\vdots & \vdots & \ddots & \vdots & \vdots\\
0 & 0 & \cdots & 1 & 0
\end{array}
\right),
\end{equation}
where $\omega=e^{i2\pi /d}$. Note that
$AB=\omega BA$ \cite{Massar2008}. Now let us discuss the following two situations:

\emph{Case $d=3$}. The pure state is $|\psi\rangle=\frac{\sqrt{2}}{2}\cos\theta|0\rangle+\frac{\sqrt{2}}{2}\cos\theta|1\rangle-\sin\theta|2\rangle$,
the unitary operators are
\begin{equation}
A=\left(
\begin{array}{ccc}
1 & 0 & 0 \\
0 & e^{\frac{2\pi i}{3}} & 0 \\
0 & 0 & e^{\frac{4\pi i}{3}}
\end{array}
\right),
B=\left(
\begin{array}{ccc}
0 & 0 & 1 \\
1 & 0 & 0 \\
0 & 1 & 0
\end{array}
\right).
\end{equation}

Then the lower bounds $S_{21}, S_{31}, S_{32}$ can be computed, it is readily obtained that $\Delta A^{2}\Delta B^{2}\geq S_{21}=I_{2}\geq S_{31}\geq S_{32}=I_{3}$. From the subgraph of Fig 4. we can clearly see that a black dotted curve $S_{31}$ be added between $I_{2}$ and $I_{3}$. This means that this descending sequence is refined.

\emph{Case $d=4$}. The pure state is $|\psi\rangle=\frac{\sqrt{3}}{3}\cos\theta|0\rangle+\frac{\sqrt{3}}{3}\cos\theta|1\rangle
+\frac{\sqrt{3}}{3}\cos\theta|2\rangle-\sin\theta|3\rangle$. $A$ and $B$ are the unitary operators:
\begin{equation}
A=\left(
\begin{array}{cccc}
1 & 0 & 0 & 0\\
0 & e^\frac{\pi i}{2} & 0 & 0\\
0 & 0 & e^{\pi i} & 0\\
0 & 0 & 0 & e^\frac{3\pi i}{2}
\end{array}
\right),
B=\left(
\begin{array}{cccc}
0 & 0 & 0 & 1\\
1 & 0 & 0 & 0\\
0 & 1 & 0 & 0\\
0 & 0 & 1 & 0
\end{array}
\right).
\end{equation}

We can calculate that $\Delta A^{2}\Delta B^{2}\geq S_{21}=I_{2}\geq S_{31}\geq S_{32}=I_{3}\geq S_{41}\geq S_{42}\geq S_{43}=I_{4}$, the three dotted curves in Fig.5 represent the three terms $S_{31}, S_{41}$ and $S_{42}$ respectively. From the subgraph of Fig.5, we can clearly see the curve of the items we added, which makes the original descending sequence more dense.

\begin{figure}[h]
\begin{center}
\centerline{\includegraphics[width=0.6\textwidth]{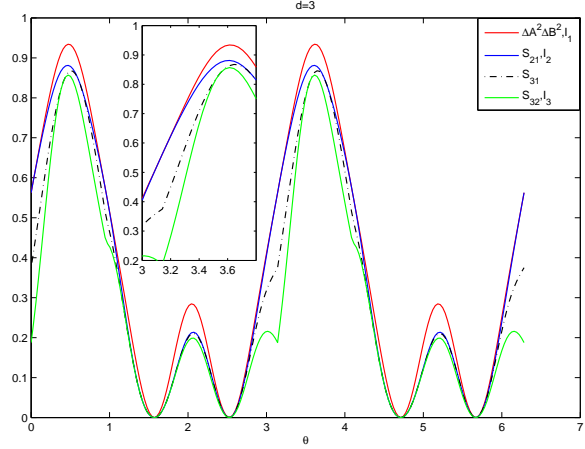}}
\caption{\textbf{Comparison of our bound with Yu \emph{\textbf{et al}}.'s bound for pure state.} The solid red, blue and green curves represent Yu \etal.'s bounds $\Delta A^2\Delta B^2(I_{1})$, $I_{2}$ and $I_{3}$ respectively.
   The black dotted curve represents $S_{31}$.}
\label{fig4}
\end{center}
\end{figure}

\begin{figure}[h]
\begin{center}
\centerline{\includegraphics[width=0.6\textwidth]%
  {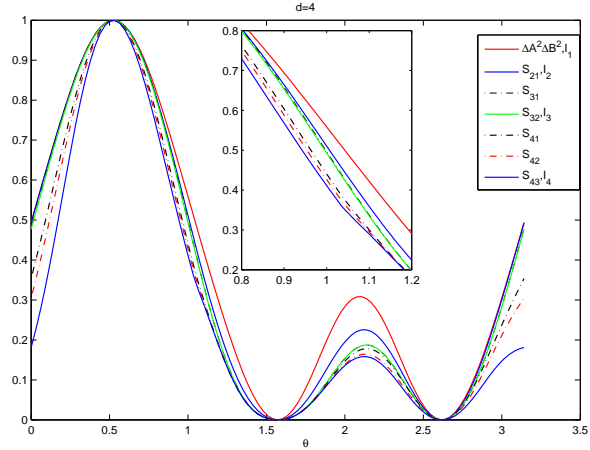}}
\caption{\textbf{Comparison of our bound with Yu \emph{\textbf{et al}}.'s bound for pure state.} The four solid curves represent $\Delta A^2\Delta B^2$, $S_{21}$, $S_{32}$ and $S_{43}$ respectively. The three dotted curves represent $S_{31}, S_{41}$  and  $S_{42}$.}
\label{fig5}
\end{center}
\end{figure}

\section{Uncertainty relations for three unitary operators}

From Theorem 1, we obtained the product-form variance-based unitary uncertainty relations of two unitary operators. Next we will study the unitary uncertainty relations of three unitary operators based on the quantities $S_{pq}$. Here in order not to cause ambiguity, we mark $S_{pq}$ as $S_{pq}^{(xy)}$, $S_{pq}^{(yz)}$ and $S_{pq}^{(xz)}$ by coordinates which represent polynomials about the variables $xy$, $yz$ and $xz$ respectively.

Let $A$, $B$ and $C$  be three unitary operators defined on an $n$-dimensional Hilbert space. Similarly suppose $\Delta A^{2}=|\vec{X}|^2$, $\Delta B^{2}=|\vec{Y}|^2$, $\Delta C^{2}=|\vec{Z}|^2$ for the (nonnegative) real vectors $\vec{X}=(x_1, x_2, \cdots, x_n)$, $\vec{Y}=(y_1, y_2, \cdots, y_n)$, $\vec{Z}=(z_1, z_2, \cdots, z_n)$, where $ x_i=|\alpha_i|$, $y_j=|\beta_j|$, $z_k=|\gamma_{k}|$.
Thus we have $\Delta A^2\Delta B^2\Delta C^{2}=|\vec{X}|^2|\vec{Y}|^2|\vec{Z}|^2
=\sum\limits_{i,j,k}^{n}x_i^2y_j^2z_{k}^{2}$.

For three unitary operators, we first propose a variable, and establish the relationship of descending sequence between two unitary operators and three unitary operators. Hence we can apply the improved decreasing sequence of the two unitary operators to derive the lower bound of uncertainty relations of the three unitary operators.

Based on the results of two unitary operators, each term in the next sum is successively scaled by using Cauchy-Schwarz inequality. We can get
\begin{eqnarray}
\sum\limits_{i,j,k}^{n}x_i^2y_j^2z_{k}^{2}&=z_{1}^{2}\sum_{i,j}^{n}x_{i}^{2}y_{j}^{2} + z_{2}^{2}\sum_{i,j}^{n}x_{i}^{2}y_{j}^{2} + \cdots + z_{n}^{2}\sum_{i,j}^{n}x_{i}^{2}y_{j}^{2}\nonumber\\
&\geq z_{1}^{2}S_{n(n-1)}^{(xy)}+z_{2}^{2}\sum_{i,j}^{n}x_{i}^{2}y_{j}^{2} + \cdots + z_{n}^{2}\sum_{i,j}^{n}x_{i}^{2}y_{j}^{2}\nonumber\\
&\geq z_{1}^{2}S_{n(n-1)}^{(xy)}+z_{2}^{2}S_{n(n-1)}^{(xy)} + \cdots + z_{n}^{2}\sum_{i,j}^{n}x_{i}^{2}y_{j}^{2}\nonumber\\
&\geq \cdots\nonumber\\
&\geq z_{1}^{2}S_{n(n-1)}^{(xy)}+z_{2}^{2}S_{n(n-1)}^{(xy)} + \cdots + z_{n}^{2}S_{n(n-1)}^{(xy)}\nonumber\\
&=(z_{1}^{2}+z_{2}^{2}+\cdots+z_{n}^{2})S_{n(n-1)}^{(xy)},
\end{eqnarray}
where the equality holds if and only if $x_iy_j=x_jy_i$ for all $i\neq j$.

Similarly, we refine the inequality by introducing a sequence of partial ones. For each $t=1,2,\cdots,n$, $p=1,2,\cdots,n$, $q=1,2,\cdots,(p-1)$, we define
\begin{equation}
M_{tpq}^{(z)}=M_{010}^{(z)}+\sum_{r=1}^{t-1}z_{r}^{2}(S_{n(n-1)}^{(xy)}
-S_{10}^{(xy)})+z_{t}^{2}(S_{pq}^{(xy)}-S_{10}^{(xy)}),
\end{equation}
where $M_{010}^{(z)}=\sum\limits_{i,j,k}^{n}x_i^2y_j^2z_{k}^{2}$.
Similarly, we have
\begin{eqnarray}
M_{tpq}^{(x)}=M_{010}^{(x)}+\sum\limits_{r=1}^{t-1}x_{r}^{2}(S_{n(n-1)}^{(yz)}-S_{10}^{(yz)})
+x_{t}^{2}(S_{pq}^{(yz)}-S_{10}^{(yz)}),\\
M_{tpq}^{(y)}=M_{010}^{(y)}+\sum\limits_{r=1}^{t-1}y_{r}^{2}(S_{n(n-1)}^{(xz)}-S_{10}^{(xz)})
+y_{t}^{2}(S_{pq}^{(xz)}-S_{10}^{(xz)}).
\end{eqnarray}

It is easily seen that
\begin{eqnarray*}
M_{tpq}^{(z)}-M_{tp(p-1)}^{(z)}=z_{t}^{2}(S_{pq}^{(xy)}-S_{p(q-1)}^{(xy)})\leq 0,\\
M_{tp1}^{(z)}-M_{t(p-1)(p-2)}^{(z)}=z_{t}^{2}(S_{p1}^{(xy)}-S_{(p-1)(p-2)}^{(xy)})\leq 0,\\
M_{tp1}^{(z)}-M_{(t-1)n(n-1)}^{(z)}=z_{t}^{2}(S_{p1}^{(xy)}-S_{10}^{(xy)})\leq 0.
\end{eqnarray*}
Thus we obtains the following descending sequence
\begin{eqnarray}
M_{010}^{(z)}&\geq M_{121}^{(z)}\geq M_{131}^{(z)}\geq M_{132}^{(z)}\geq \cdots \geq M_{1n(n-1)}^{(z)}\nonumber\\& \geq M_{221}^{(z)}\geq M_{231}^{(z)}\geq M_{232}^{(z)}\geq \cdots M_{2n(n-1)}^{(z)}\nonumber\\
&\geq \cdots \nonumber\\&\geq M_{n21}^{(z)}\geq M_{n31}^{(z)}\geq M_{n32}^{(z)}\geq \cdots \geq M_{nn(n-1)}^{(z)}.
\end{eqnarray}

\emph{Theorem} 2. For a fixed quantum state $\rho$ and three unitary operators $A$, $B$ and $C$ on an $n$-dimensional Hilbert space $H$, the product of the variances obeys the following inequalities ($t=1,2,\cdots,N;  p=1,2,\cdots,N;  q=1,2,\cdots,(p-1)$)
\begin{equation}
\Delta A^2\Delta B^2\Delta C^2\geq \max\{M_{tpq}^{(x)},M_{tpq}^{(y)},M_{tpq}^{(z)}\}=M_{tpq},
\end{equation}
where $N=n$ (or $n^2$) if $\rho$ is pure (or mixed).

\emph{Proof}. According to the quantities $M_{010}^{(z)}$ given above, we have $M_{010}^{(z)}=\sum\limits_{i,j,k}^{n}x_{i}^{2}y_{j}^{2}z_{k}^{2}=\Delta A^{2}\Delta B^{2}\Delta C^{2}$. Then from the recursive relation given above, the following recursive terms can be obtained
\begin{eqnarray*}
M_{121}^{(z)}-M_{010}^{(z)}=z_{1}^{2}(S_{21}^{(xy)}-S_{10}^{(xy)})\leq 0,\\
M_{131}^{(z)}-M_{121}^{(z)}=z_{1}^{2}(S_{31}^{(xy)}-S_{21}^{(xy)})\leq 0,\\
M_{132}^{(z)}-M_{131}^{(z)}=z_{1}^{2}(S_{32}^{(xy)}-S_{31}^{(xy)})\leq 0,\\
\qquad \vdots\\
M_{1n(n-1)}^{(z)}-M_{1n(n-2)}^{(z)}=z_{1}^{2}(S_{n(n-1)}^{(xy)}-S_{10}^{(xy)})\leq 0,\\
M_{221}^{(z)}-M_{1n(n-1)}^{(z)}=z_{2}^{2}(S_{21}^{(xy)}-S_{10}^{(xy)})\leq 0,\\
M_{231}^{(z)}-M_{221}^{(z)}=z_{2}^{2}(S_{31}^{(xy)}-S_{21}^{(xy)})\leq 0,\\
\qquad \vdots\\
M_{2n(n-1)}^{(z)}-M_{2n(n-2)}^{(z)}=z_{2}^{2}(S_{n(n-1)}^{(xy)}-S_{n(n-2)}^{(xy)})\leq 0,\\
\qquad \vdots\\
M_{n21}^{(z)}-M_{(n-1)n(n-1)}^{(z)}=z_{n}^{2}(S_{21}^{(xy)}-S_{10}^{(xy)})\leq 0,\\
M_{n31}^{(z)}-M_{n21}^{(z)}=z_{n}^{2}(S_{31}^{(xy)}-S_{21}^{(xy)})\leq 0,\\
\qquad \vdots\\
M_{nn(n-1)}^{(z)}-M_{nn(n-2)}^{(z)}=z_{n}^{2}(S_{n(n-1)}^{(xy)}-S_{n(n-2)}^{(xy)})\leq 0.
\end{eqnarray*}
Therefore we obtain the decreasing sequence
\begin{eqnarray*}
M_{010}^{(z)}&\geq M_{121}^{(z)}\geq M_{131}^{(z)}\geq M_{132}^{(z)}\geq \cdots \geq M_{1n(n-1)}^{(z)}\\
&\geq M_{221}^{(z)}\geq M_{231}^{(z)}\geq M_{232}^{(z)}\geq \cdots M_{2n(n-1)}^{(z)}\\
&\geq \cdots \\
&\geq M_{n21}^{(z)}\geq M_{n31}^{(z)}\geq M_{n32}^{(z)}\geq \cdots \geq M_{nn(n-1)}^{(z)}.
\end{eqnarray*}
Then we have
\begin{eqnarray*}
\Delta A^{2}\Delta B^{2}\Delta C^{2}=M_{010}^{(z)}\geq M_{121}^{(z)}\geq\cdots\geq M_{tpq}^{(z)}\geq\cdots\geq M_{nn(n-1)}^{(z)}.
\end{eqnarray*}
Hence we prove that $\Delta A^{2}\Delta B^{2}\Delta C^{2}\geq M_{tpq}^{(z)}$. Similarly, we can get that $\Delta A^{2}\Delta B^{2}\Delta C^{2}\geq M_{tpq}^{(x)}$, $\Delta A^{2}\Delta B^{2}\Delta C^{2}\geq M_{tpq}^{(y)}$. Thus
we have $\Delta A^2\Delta B^2\Delta C^2\geq \max\{M_{tpq}^{(x)},M_{tpq}^{(y)},M_{tpq}^{(z)}\}=M_{tpq}$.

For two and three unitary operators, similarly, we can define
\begin{eqnarray}
(\pi_{1}, \pi_{2})S_{pq}^{(xy)}=&-\sum\limits_{j=2\atop j>i}^{p-1}\sum\limits_{i=1}^{j-1}(x_{\pi_{1}(j)}y_{\pi_{2}(i)}
-x_{\pi_{2}(i)}y_{\pi_{1}(j)})^{2}\nonumber\\
&-\sum\limits_{m=1}^{q}(x_{\pi_{1}(p)}y_{\pi_{2}(m)}
-x_{\pi_{2}(m)}y_{\pi_{1}(p)})^{2}\nonumber\\
&+\sum\limits_{i,j}^{n}x_{\pi_{1}(i)}^{2}y_{\pi_{2}(j)}^{2}.
\end{eqnarray}

Meanwhile, we obtain $(\pi_{1}, \pi_{2})S_{pq}^{(yz)}$ and $(\pi_{1}, \pi_{2})S_{pq}^{(xz)}$. Then
\begin{equation}
\Delta A^{2}\Delta B^{2}\geq \max\limits_{\pi_{1}, \pi_{2}\in S_{N}}\{\max\{(\pi_{1}, \pi_{2})S_{pq}^{(xy)},(\pi_{1}, \pi_{2})S_{pq}^{(yz)},
(\pi_{1}, \pi_{2})S_{pq}^{(xz)}\}\}.
\end{equation}

Similarly, we can define $(\pi_{1}, \pi_{2},\pi_{3})M_{tpq}^{(x)}$,  $(\pi_{1}, \pi_{2},\pi_{3})M_{tpq}^{(y)}$, $(\pi_{1}, \pi_{2},\pi_{3})M_{tpq}^{(z)}$, then
\begin{eqnarray}
\Delta A^{2}\Delta B^{2}\Delta C^{2}\geq \max\limits_{\pi_{1}, \pi_{2},\pi_{3}\in S_{N}}\{\max\{&(\pi_{1}, \pi_{2},\pi_{3})M_{tpq}^{(x)},
(\pi_{1}, \pi_{2},\pi_{3})M_{tpq}^{(y)},\nonumber\\&(\pi_{1}, \pi_{2},\pi_{3})M_{tpq}^{(z)}\}\}.
\end{eqnarray}

\emph{Example} 4. Suppose $|\varphi\rangle=\frac{\sqrt{2}}{2}\cos\frac{\theta}{2}|0\rangle+
\frac{\sqrt{2}}{2}\sin\frac{\theta}{2}|1\rangle-\sin\frac{\theta}{2}|2\rangle$ is the pure state.
$A$, $B$ and $C$ are three unitary operators, which can be denoted as follows:
\begin{equation}
A=\left(
\begin{array}{ccc}
1 & 0 & 0 \\
0 & e^\frac{\pi i}{2} & 0 \\
0 & 0 & e^\frac{3\pi i}{2}
\end{array}
\right),
B=\left(
\begin{array}{ccc}
0 & 0 & 1 \\
1 & 0 & 0 \\
0 & 1 & 0
\end{array}
\right),
C=\left(
\begin{array}{ccc}
0 & 1 & 0 \\
1 & 0 & 0 \\
0 & 0 & 1
\end{array}
\right).
\end{equation}

Using Theorem 2, the lower bound $M_{tpq}^{(z)}$ can be easily calculated, we have
\begin{eqnarray*}
&\Delta A^{2}\Delta B^{2}\Delta C^{2}\geq M_{121}^{(z)}\geq M_{131}^{(z)}\geq M_{132}^{(z)}\\
&\geq M_{221}^{(z)}\geq M_{231}^{(z)}\geq M_{232}^{(z)}\geq M_{321}^{(z)}\geq M_{331}^{(z)}\geq M_{332}^{(z)}.
\end{eqnarray*}
Fig.6 shows that these lower bound curves $M_{121}^{(z)}$, $M_{131}^{(z)}$, $M_{132}^{(z)}$ and the bound $(I_{1}J_{1}K_{1})^{\frac{1}{2}}$, $(I_{2}J_{2}K_{2})^{\frac{1}{2}}$ in \cite{Yu2019}. As shown in the figure, we obtain the lower bound is tighter than the bound \cite{Yu2019} for three unitary operators. Meanwhile, we bound $M_{121}^{(z)}$ is the most tight under these circumstances. Certainly, we can obtain
\begin{eqnarray*}
&\Delta A^{2}\Delta B^{2}\Delta C^{2}\geq M_{121}\geq M_{131}\geq M_{132}\\
&\geq M_{221}\geq M_{231}\geq M_{232}\geq M_{321}\geq M_{331}\geq M_{332}.
\end{eqnarray*}
we can easily see that the lower bound is tighter after seeking the maximum value from Fig.7. In order to make our observations clearly, we draw a partial lower bound curves in Fig.7, which is also enough to show that our bound is tighter.

\begin{figure}[h]
\begin{center}
\centerline{\includegraphics[width=0.6\textwidth]%
  {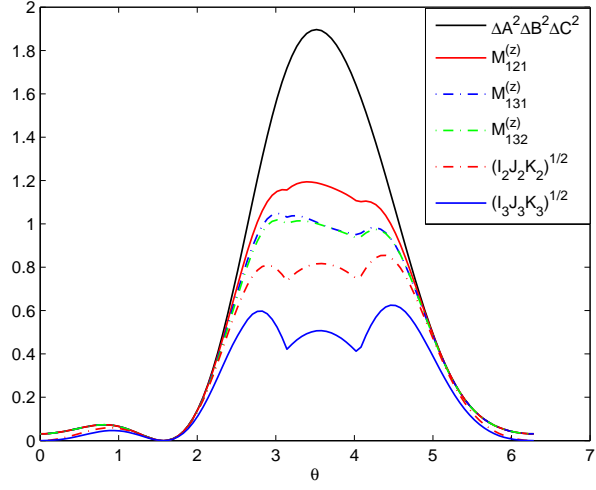}}
\caption{\textbf{Comparison of our bounds with Yu \emph{\textbf{et al}}.'s for pure state.} The solid black (upper) curve is $\Delta A^2\Delta B^2\Delta C^2$, the solid red curves and the two dotted blue, dotted green represent our bounds $M_{121}^{(z)}$,  $M_{131}^{(z)}$, $M_{132}^{(z)}$ respectively. The dotted red curves and solid blue curves represent Yu et al.'s bounds $(I_{2}J_{2}K_{2})^{\frac{1}{2}}$ and $(I_{3}J_{3}K_{3})^{\frac{1}{2}}$.}
\label{fig6}
\end{center}
\end{figure}

\begin{figure}
\begin{center}
\centerline{\includegraphics[width=0.6\textwidth]%
  {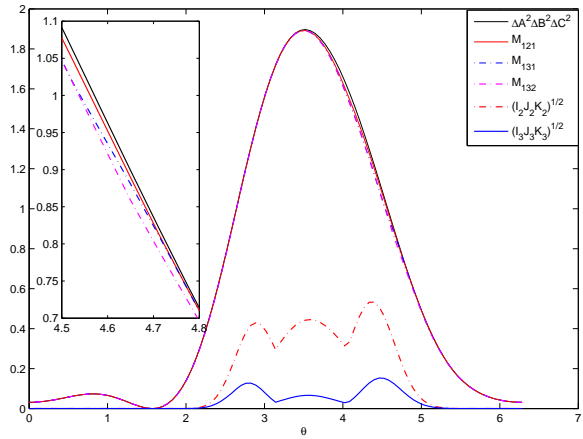}}
\caption{\textbf{Comparison of our bounds with Yu \emph{\textbf{et al}}.'s bound for pure state.} The curves $M_{121}=\max\{M_{121}^{(x)},M_{121}^{(y)},M_{121}^{(z)}\}$, $M_{131}=\max\{M_{131}^{(x)},M_{131}^{(y)},M_{131}^{(z)}\}$, and $M_{132}=\max\{M_{132}^{(x)},M_{132}^{(y)},M_{132}^{(z)}\}$ , then the dotted red curves and solid blue curves represent Yu et al.'s bounds $(I_{2}J_{2}K_{2})^{\frac{1}{2}}$ and $(I_{3}J_{3}K_{3})^{\frac{1}{2}}$, respectively.}
\label{fig7}
\end{center}
\end{figure}

\section{Conclusion}

In this paper, we present the lower bound of unitary uncertainty relations for two unitary operators and a class of states. Meanwhile, we improved the descending sequence in \cite{Yu2019} to be finer. For three unitary operators, we got a sequence of lower bounds
$M_{tpq}$ by using the improved descending sequence. Moreover, our bounds
$M_{tpq}$ is tighter than the lower bound in \cite{Yu2019}. In the article, we only researched the lower bound of uncertainty relations with two and three unitary operators. Certainly, this method can also be generalized to the boundary problem of multiple unitary operators.

\section*{Acknowledgments}
This work was supported by the Hebei Provincial Natural Science Foundation of China (Grant No.A2019210057) and the National Natural Science Foundation of China under Grant No.11847105.

\section*{Reference}

\bibliographystyle{amsalpha}

\end{document}